\documentclass[aps,pre,showpacs,onecolumn,floatfix,superscriptaddress]{revtex4}

\usepackage{amsmath} \usepackage{amsfonts} \usepackage{amssymb}
\usepackage{graphicx}
\usepackage{dcolumn}
\usepackage{bm}
\usepackage{color}
\usepackage{epstopdf}

\newcommand{\beq}{\begin{equation}}
\newcommand{\eeq}{\end{equation}}
\newcommand{\eps}{\epsilon}

\begin{document}

\title{Interaction of Traveling Waves with
Mass-With-Mass Defects within a Hertzian Chain}

\author{P.G. Kevrekidis}
\affiliation{Department of Mathematics and Statistics, University
of Massachusetts, Amherst, Massachusetts 01003-4515, USA}

\author{A. Vainchtein}
\affiliation{Department of Mathematics, University of Pittsburgh, Pittsburgh, Pennsylvania 15260, USA}

\author{M. Serra Garcia}
\affiliation{Graduate Aeronautical Laboratories (GALCIT) and Department of Applied Physics, California Institute of Technology, Pasadena, California 91125, USA}

\author{C. Daraio}
\affiliation{Graduate Aeronautical Laboratories (GALCIT) and Department of Applied Physics, California Institute of Technology, Pasadena, California 91125, USA}

\date{\today}

\begin{abstract}
We study the dynamic response of a granular chain of particles with a resonant inclusion (i.e., a particle attached to a harmonic oscillator, or a mass-with-mass defect).
We focus on the response of granular chains excited by an impulse, with no static precompression. We find that the presence of the harmonic oscillator can be used to tune the transmitted and reflected energy of a mechanical pulse by adjusting the ratio between the harmonic resonator mass and the bead mass. Furthermore, we find that  this system has the capability of asymptotically trapping energy, a feature that is not present in granular chains containing other types of defects. Finally, we study the limits of low and high resonator mass, and the structure of the reflected and transmitted pulses.
\end{abstract}


\maketitle

\section{Introduction}
\label{sec:intro}

Granular crystals consisting of tightly packed arrays of solid particles, elastically deforming upon contact, present interesting dynamical features that enabled fundamental physical discoveries and suggested new engineering applications. Most studies of granular crystals focused on the highly nonlinear dynamic response of these systems ~\cite{nesterenko1,sen08,nesterenko2,coste97,dar06,hong05,fernando,doney06,dev08,dar05,spadoni10,dar05b,ahnert,linden,hascoet,job,vakakis,theo,chiaradef,chiara_nature,theocharis_pre,sen}. One-dimensional (1D) granular crystals have been studied in a number of analytical, numerical and experimental investigations; see in~\cite{sen08} for a recent review of this topic. The ability to use a wide variety of materials and bead sizes, as well as the tunability of the response within the weakly or strongly nonlinear regime makes granular crystals a natural paradigm for physical explorations of wave phenomena and the effect of nonlinearity on them~\cite{nesterenko2,coste97}. On the engineering side, this tunability makes such crystals promising candidates for numerous applications, including shock and energy absorbing materials \cite{dar06,hong05,fernando,doney06}, actuating and focusing devices \cite{dev08, spadoni10}, and sound scramblers or filters \cite{dar05,dar05b,chiara_nature}.

To model the dynamics of granular chains, Hertzian interparticle interactions (proportional to the relative displacement of adjacent bead centers raised to the $3/2$ power in the case
of spherical beads) have been established as the canonical approach~\cite{nesterenko1}. These chains have been shown to support the emergence of nonlinear traveling waves, which have been described through different types of partial differential equation models (see, e.g.,~\cite{ahnert}) or even by binary collision particle models (see, e.g.,~\cite{linden}). Although these waves are treated as compactly supported in the continuum approximations, they decay with a doubly exponential power law~\cite{english,atanas} (i.e. extremely fast, but their support is not genuinely compact).

Once a chain of particles is excited by an impulse, more than 99\%  of its energy is rapidly and spontaneously rearranged into one or more of such nonlinear traveling waves (TW)~\cite{hinch}. Examining the interaction of the resulting TWs with a defect has been of particular interest from the point of view of applications (e.g. for detecting cracks \cite{dev08} or for detecting other irregularities in a medium (see e.g., \cite{sen_etal05} and references therein). A pioneering study in this regard was the computational work of~\cite{hascoet}, which examined both the case of a light defect and the resulting symmetric emission of traveling waves in both directions by the defect, and that of a heavy defect, which produced a train of solitary waves asymmetrically to the right of the defect. The theme of impact upon a light defect has been revisited in the work of~\cite{job} where the synergy of experiments, numerical computations and analytical approximations demonstrated the possibility of transient breather formation in the system. Very recently, the problem was also revisited from a chiefly analytical point of view in~\cite{vakakis}, where a reduction of the problem to a chain of three beads solved using a multi-scale expansion was used. This enabled an 
accurate capturing of the slow dynamics of the defect bead and its neighbors, along with the fast transient breathing dynamics of the light bead. It should be noted that defects have also been studied in such granular chains in the presence of a precompression force (creating an underlying linear limit and hence the potential for a weakly nonlinear regime) in the bulk, where the formation of defect breathing modes has been elaborated both analytically/numerically~\cite{theo} and numerically/experimentally~\cite{chiaradef}. The usefulness of defect breathing modes (at one edge of the chain) as generators of acoustic diode type effects has also been explored recently in~\cite{chiara_nature}.
We also mention in passing the consideration of interactions of
traveling breathers with defects in a ``Newton's cradle'' model
where the Hertzian chain has a local oscillator associated
with each bead~\cite{james1}.

\begin{figure}[tbph]
\begin{center}
\includegraphics[width=3.0in]{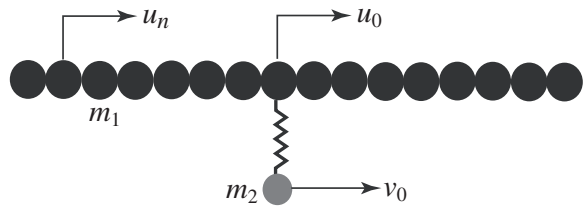}
\caption{Diagram of a granular chain containing a mass-with-mass defect.}
\label{fig:MwM}
\end{center}
\end{figure}

\begin{figure}[tbph]
\begin{center}
\includegraphics[width=5.0in]{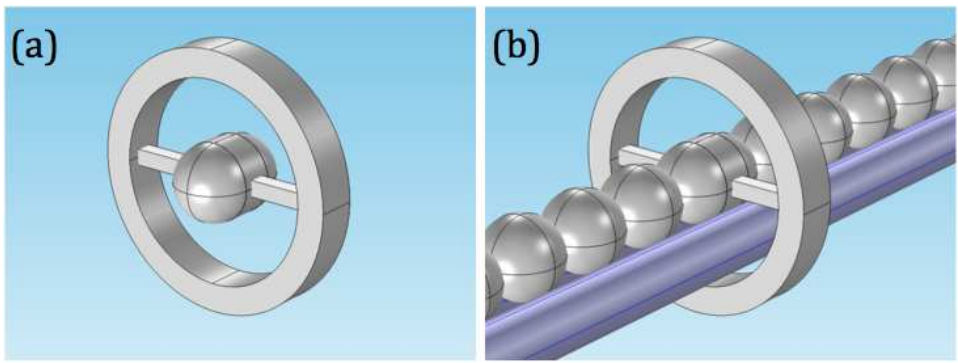}
\caption{Suggested experimental implementation of the mass-with-mass defect. (a) Defect bead with the ring resonator that provides the mass-with-mass. (b) Part of the suggested experimental setup containing the MwM defect.}
\label{fig:MwMexp}
\end{center}
\end{figure}

In this paper, we will consider the interaction of a solitary wave with a new kind of defect, consisting of a secondary mass attached to one of the beads of the chain through a linear spring, as shown in Fig.~\ref{fig:MwM}.
We will refer to this defect as a mass with mass (MwM) defect.
A MwM defect can be implemented experimentally by attaching a resonant solid structure to one of the beads, as shown in Fig.~\ref{fig:MwMexp}. Such a structure should be designed carefully to avoid higher-order normal modes of the resonant structure to participate in the dynamics of the system. A ring resonator vibrating in its piston normal mode would satisfy this requirement and is able to provide the values of spring stiffness and secondary mass
needed to reproduce experimentally the effects that we have predicted theoretically.

Meanwhile, the interaction of the defect bead with the neighboring particles is preserved as Hertzian.
A feature of a granular chain with a MwM defect
is the existence of an underlying linear oscillator at the defect. This oscillator presents the potential for
long-term energy trapping, a feature that cannot be present in the mass defects considered in earlier
studies. We examine various limits of the system, especially focusing
on the limit of small secondary mass, which is analytically tractable and able to reproduce much of the physics
of the interaction between a solitary wave and a MwM defect.
We also separately consider
the limit of very large secondary mass, where the reflected pulse has much larger energy, as well as the intermediate
regime, where an emission of a train of transmitted solitary waves that have
progressively decreasing amplitudes and speeds is observed.

Our presentation is organized as follows. In Section~\ref{sec:model}, we introduce the model and the
associated quantities that will be monitored. In Section~\ref{sec:numer},
we discuss the numerical observations of a traveling wave interacting with the MwM defect
in the regimes of a small, intermediate and large secondary mass. We also briefly comment on the
case of several defects. In Section~\ref{sec:reduced} we analyze the limit of small secondary mass using two approaches: a direct perturbation method that approximates the system in this limit as a local oscillator driven by a solitary wave
and a multiscale analysis of a reduced model that captures some of the prototypical features of the full system.
Section~\ref{sec:conclusions} contains the summary of our findings and the discussion of possible future directions.

\section{The model}
\label{sec:model}
We will consider a granular chain of identical spherical elastic beads of mass $m_1$ each. Let $u_n(t)$ denote the
displacement of the $n$th bead from its equilibrium position and denote $\dot{u}_n(t)=u_n'(t)$, $\ddot{u}_n(t)=u_n''(t)$. The interaction between $n$th and $(n+1)$th beads is governed by
Hertz interaction potential
\[
U=\dfrac{2a}{5}(u_n-u_{n+1})_+^{5/2},
\]
where $(x)_+=x$ when $x>0$ and equals zero otherwise, and $a>0$ is a material constant.
The beads thus interact only when they overlap and are not subject to a force when the overlap
is absent.
The defect bead, at $n=0$, is attached to another mass, $m_2$, via a linear spring of stiffness $K>0$, as shown in Fig.~\ref{fig:MwM}. This mass is constrained to move in the horizontal direction, with displacement $v_0(t)$. The equations of motion are:
\beq
\begin{split}
m_1\ddot{u}_n&=a\biggl((u_{n-1}-u_n)_+^{3/2}-(u_{n}-u_{n+1})_+^{3/2}\biggr)-K(u_0-v_0)\delta_{n0}\\
m_2\ddot{v}_0&=K(u_0-v_0),
\end{split}
\label{eq:orig_eqns}
\eeq
where we used the Kronecker delta, $\delta_{n0}=1$ when $n=0$ and zero otherwise.
We assume that all masses, except the $j$th bead, for some $j<0$, are initially at rest, and the beads in the chain just touch their neighbors at $t=0$. The $j$th bead is excited by setting its initial velocity to $V>0$. The initial
conditions are thus:
\beq
u_n(0)=v_0(0)=0, \qquad \dot{v}_0(0)=0=\dot{u}_n(0), \quad n \neq j, \qquad \dot{u}_j(0)=V, \quad j<0.
\label{eq:orig_ICs}
\eeq

It is convenient to rescale \eqref{eq:orig_eqns}, \eqref{eq:orig_ICs} by introducing dimensionless displacements $\bar{u}_n$, $\bar{v}_0$ and time $\bar{t}$ related to the original variables via \cite{linden}
\beq
u_n=\biggl(\dfrac{m_1V^2}{a}\biggr)^{2/5}\bar{u}_n, \quad v_0=\biggl(\dfrac{m_1V^2}{a}\biggr)^{2/5}\bar{v}_0,
\quad
t=\dfrac{1}{V}\biggl(\dfrac{m_1V^2}{a}\biggr)^{2/5}\bar{t}.
\label{eq:rescaling}
\eeq
The two dimensionless parameters are
\beq
\eps=\dfrac{m_2}{m_1},
\label{eq:eps}
\eeq
the ratio of the two masses,
and
\beq
\kappa=\dfrac{K}{m_1^{1/5}a^{4/5}V^{2/5}},
\label{scaling}
\eeq
which measures the strength of the linear elastic spring in the mass-with-mass defect relative to the Hertzian potential stiffness at the particular $V$.
In what follows, we set $\kappa=1$ for simplicity (for any $K$, we can
suitably select $V$ to achieve this in Eq.~(\ref{scaling})), while focusing on the effect of $\eps$. Substituting \eqref{eq:rescaling}, \eqref{eq:eps} and $\kappa=1$ into \eqref{eq:orig_eqns} and \eqref{eq:orig_ICs} and dropping the bars on the new variables, we obtain
\beq
\begin{split}
\ddot{u}_n &= (u_{n-1}-u_n)^{3/2}_+ - (u_n - u_{n+1})^{3/2}_+ - (u_0-v_0) \delta_{n0}
\\
\eps\ddot{v}_0 &= u_0 - v_0
\end{split}
\label{eqn1}
\eeq
and
\beq
u_n(0)=v_0(0)=0, \qquad \dot{v}_0(0)=0=\dot{u}_n(0), \quad n \neq j, \qquad \dot{u}_j(0)=1, \quad j<0.
\label{eq:rescaled_ICs}
\eeq

We conduct a series of numerical experiments in
the spirit of~\cite{hascoet}, in order to understand the dynamics
of the granular chain in the presence of a mass-with-mass defect.
Notice that in addition to the displacement fields
$u_n$ and $v_0$, and the corresponding velocity fields $\dot{u}_n$
and $\dot{v}_0$, another characteristic quantity the system is the total energy $E=\sum_{n} e_n$, where
\beq
e_n= \frac{1}{2} \dot{u}_n^2 + \frac{\eps}{2} \dot{v}_n^2 \delta_{n0}
+ \frac{1}{2} (u_0-v_0)^2 \delta_{n0} + \frac{1}{5} \left[
(u_{n-1}-u_n)^{5/2}_+ + (u_n - u_{n+1})^{5/2}_+ \right]
\label{end}
\eeq
is the energy density (energy stored in each bead).
The total energy $E$ is a conserved quantity of the
system. We have confirmed that in our dynamical simulations (explicit fourth order Runge-Kutta with fixed time step of $10^{-3}$) energy is conserved up
to the order of $10^{-11}$. Once the traveling wave interacts with the defect, part of the energy will be reflected,
part of the energy will be transmitted, and part of the energy will be trapped. We define the fraction of energy that is \emph{reflected} as
\beq
R=\frac{1}{E}\sum_{n<-1} e_n,
\label{eq:R}
\eeq
the fraction which is \emph{transmitted} as
\beq
T = \frac{1}{E}\sum_{n>1} e_n,
\label{eq:T}
\eeq
so that the \emph{trapped} fraction of the energy is given by
\beq
1-T-R=\frac{1}{E}\sum_{n=-1}^1 e_n.
\label{eq:trapped_en}
\eeq

We now present a detailed discussion of our numerical results and some analytic approximations of the system.

\section{Numerical Results}
\label{sec:numer}
We numerically integrate \eqref{eqn1} for $-100 \leq n \leq 100$ subject to \eqref{eq:rescaled_ICs}, with the defect located at $n=0$, in the middle of the chain. The initial excitation is at $n=j=-50$, which enables the robust formation of the traveling wave well before its impact with
the mass-with-mass defect site and precludes any backscattering or
rebounding waves from the excited site to affect the defect location
within the duration of our numerical simulations.

\begin{figure}[tbph]
\begin{center}
\includegraphics[width=12.cm]{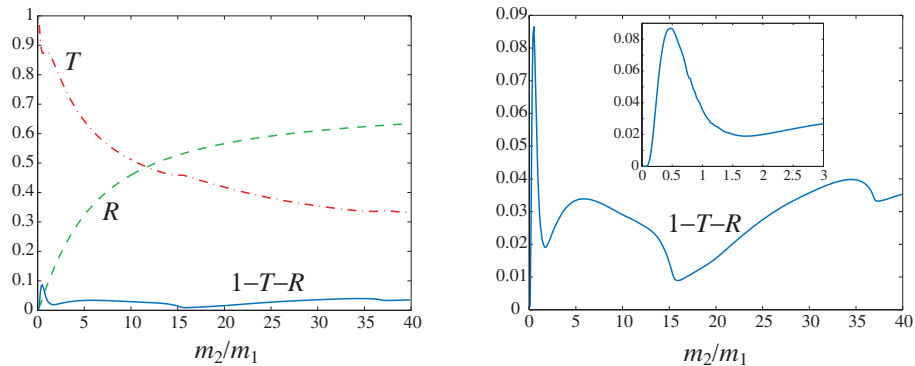}
\caption{Left panel: the different fractions of the energy are plotted as a function
of the mass ratio $\epsilon=m_2/m_1$. Right panel: trapped portion of the energy.
The transmitted energy fraction $T$ of the right
part of the chain is presented by the red dash-dotted line, the
reflected part of the energy $R$ is given by the green dashed line, while
the non-vanishing (for $\epsilon>0$) trapped fraction of the energy
is shown by the blue solid line.}
\label{line}
\end{center}
\end{figure}

Running the simulations for a range of the mass ratio $\epsilon=m_2/m_1$ values, we compute the fractions of the energy, defined in \eqref{eq:R}, \eqref{eq:T} and \eqref{eq:trapped_en}, respectively, as functions of $\epsilon$. The results are shown in Fig.~\ref{line}.
A well-understood limiting case is that of $\epsilon=0$, when $u_0=v_0$, so that the mass-with-mass defect is effectively absent and the system features near perfect transmission: $T \approx 1$ and $R \approx 0$.
This observation takes into account the well-known feature~\cite{hinch}
that over $99$\% of the impact kinetic energy is stored within
a highly nonlinear traveling wave (originally described by Nesterenko
within his quasicontinuum theory~\cite{nesterenko1}). Another physical limit is that
of large values of $\epsilon$, $m_2 \gg m_1$, when the inertia prevents the secondary mass
from moving, so $v_0 \approx 0$. In this case the transmission is far from perfect (e.g. $T \approx 0.26$ and $R \approx 0.71$ at $\epsilon=10000$), and the trapped fraction of the energy approaches the value of about $0.03$ as 
$\epsilon$ becomes large.
Observe that in the regime of comparable bead masses, between these two limits, the trapped energy fraction exhibits oscillations.

We begin by considering the case of small $\epsilon$. As $\epsilon$ departs from zero, we
still have a single propagating traveling wave within the chain, but now it experiences
a weak reflection from the defect. We also find that some of the
energy is trapped, which is a {\it fundamental} difference from the observations
in the context of light or heavy mass defects in a Hertzian
chain~\cite{hascoet,job,vakakis}.

\begin{figure}[tbph]
\begin{center}
\includegraphics[width=\textwidth]{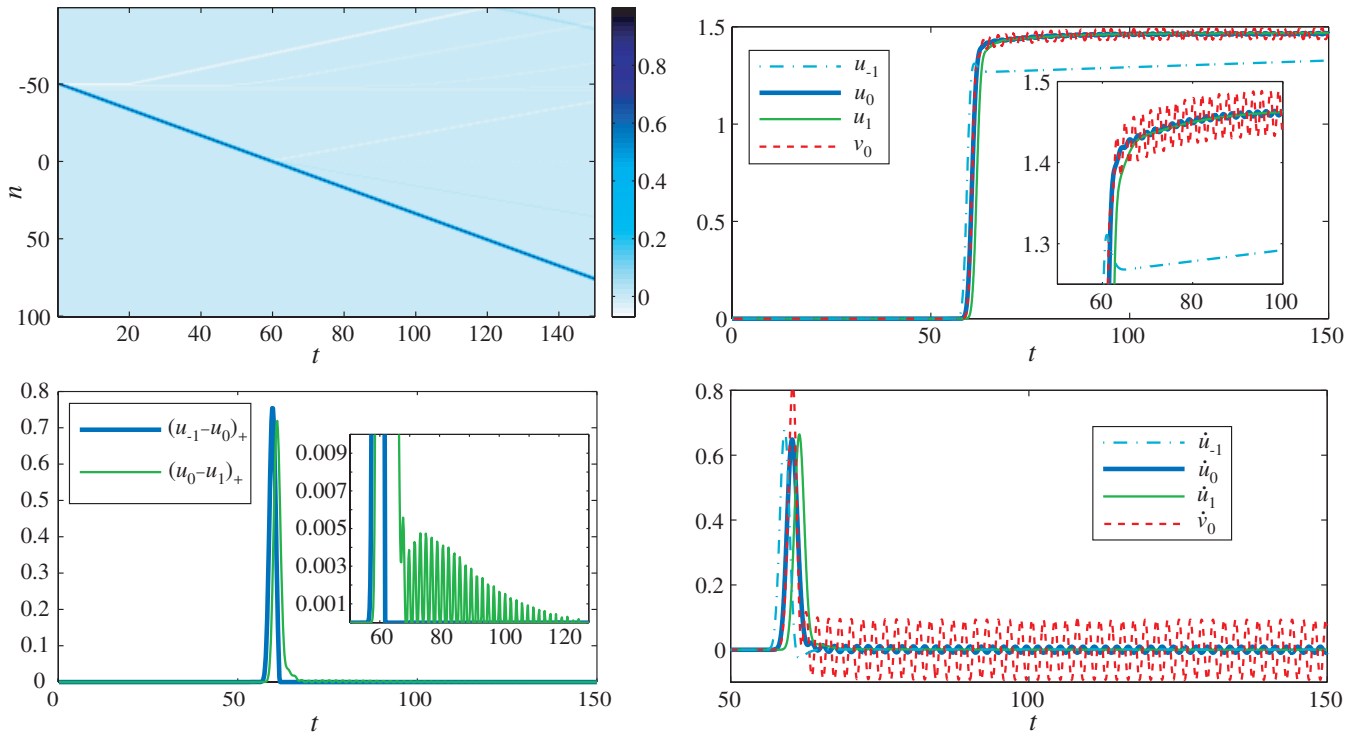}
\caption{The results of simulations at $\eps=0.1$. The top left panel presents the space-time ($n$-$t$) evolution
of the contour plot of the field $\dot{u}_n(t)$. The thick blue and green lines in the bottom left panel
show the quantities $(u_{n-1}-u_n)_+$ and $(u_n-u_{n+1})_+$, respectively, to give a sense of when
(after the impact) gap openings arise. The top right panel presents the
displacements of the central sites; cyan dash-dotted line corresponds
to $u_{-1}(t)$, thick blue solid to $u_{0}(t)$, green solid to $u_{1}(t)$
and red dashed to $v_{0}(t)$. The same pattern is followed for the corresponding
velocities at the bottom right plot.}
\label{fig01}
\end{center}
\end{figure}

This picture is corroborated by the detailed dynamics of this
case as presented in Fig.~\ref{fig01} for the value of $\epsilon=0.1$.
The contour plot in the top left panel presents the space-time evolution of the velocity
field which characteristically represents the traveling wave.
The bottom left panel below the contour plot and on the same time scale
shows the evolution of the quantities $(u_{n-1}-u_n)_+$ and $(u_n-u_{n+1})_+$, which, as we recall,
determine the forces exerted on the corresponding beads.
When the force vanishes, there is
what we call a ``gap opening''~\cite{theocharis_pre}, i.e., the beads
no longer interact. It is, thus, clear by also consulting the second
and third subplots of Fig.~\ref{fig01} that upon the impact of
the wave, there emerges a reflection and subsequently a permanent
gap opening in the interaction of beads at $n=0$ and $n=-1$.
This produces the single reflected pulse present in the process.
On the other hand, there is a more prolonged interaction between beads at
$n=0$ and $n=1$, which, in turn, leads at the early stage
of the dynamics to the emergence
of a transmitted pulse. However, a key feature also arising in the
process is the existence of a trapped part of the energy. For the
small $\eps$ considered here, this
part is weak but it is definitely present and
manifests itself as oscillations at a characteristic frequency.
Interestingly, this frequency, as we will justify in Sec.~\ref{sec:reduced}
where we analyze this asymptotic case, is precisely
the linear frequency of the mass-with-mass oscillator
$\omega=\sqrt{1+\epsilon^{-1}}$.

\begin{figure}[tbph]
\begin{center}
\includegraphics[width=\textwidth]{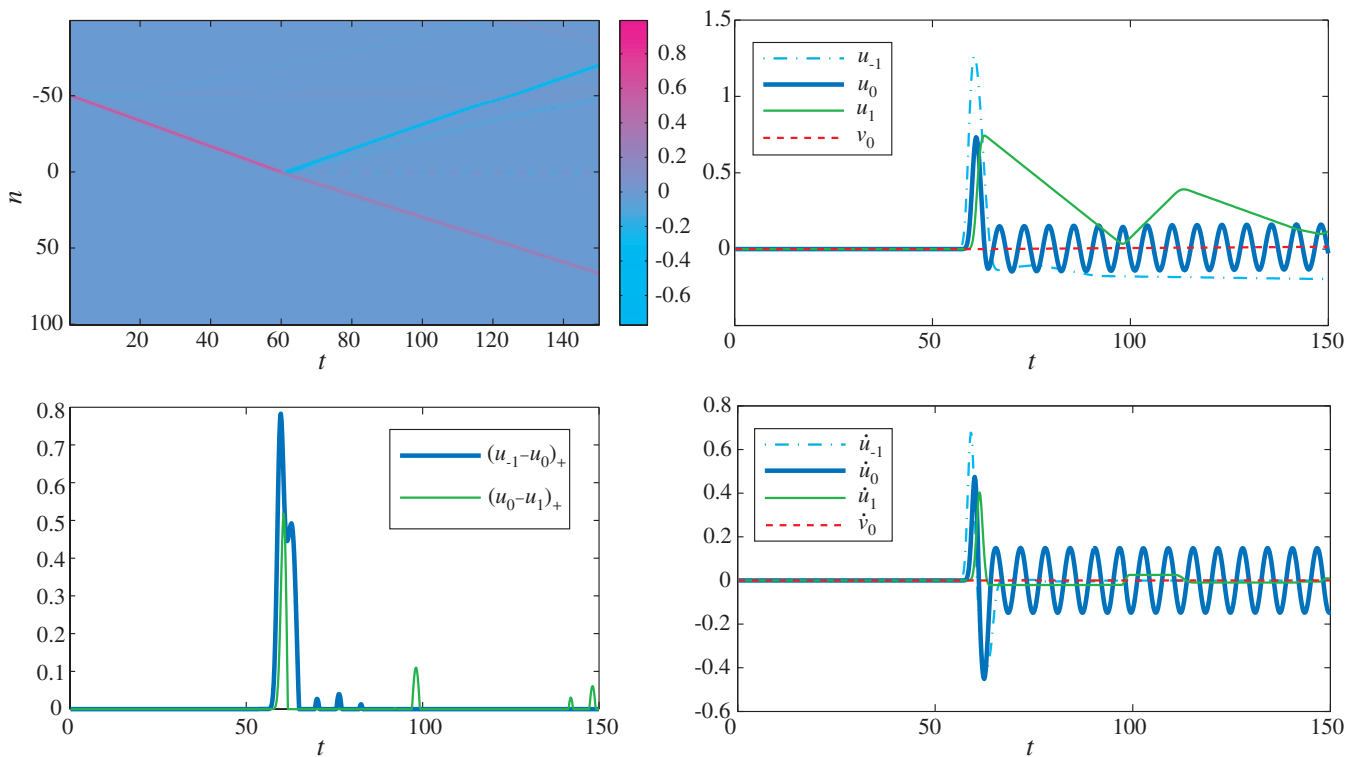}
\caption{The same properties as for the Fig.~\ref{fig01} are illustrated
but now for the case of $\epsilon=10000$.}
\label{fig10000}
\end{center}
\end{figure}

We now turn to the opposite limit, namely that of a much
larger mass $m_2$ in comparison to the mass $m_1$ of the granular
chain beads. This case is represented by Fig.~\ref{fig10000}, which
presents the computations for $\epsilon=10000$.
In this case the large secondary mass barely moves ($v_0 \approx 0$).
However, $u_0$ oscillates around the zero value after the initial solitary wave
reaches the defect site, triggering oscillations of $u_1$, whose amplitude descreases
with time. As before, the gap between $n=0$ and $n=-1$ forms, sending a reflected pulse (note, however,
the gap is no longer permanent, as these two sites 
briefly interact several times later).
As in the small $\eps$ case, there is also a transmitted pulse in this limit. However, the transmitted
pulse is much weaker in this case, while the reflected pulse is much stronger: indeed, at $\eps=10000$,
we have $R \approx 0.71$ and $T \approx 0.26$, far from the near-perfect transmission we observed at small $\eps$.
It is also important to note that there is an immediate
gap opening, upon the first passage of the wave, between $n=0$ and
$n=1$ in this case, as can be observed in Fig.~\ref{fig10000}. In this
case, the asymptotic scenario (of $\epsilon \rightarrow \infty$)
leads the central site to acquire a residual oscillation of
$u_0$ with approximately 
unit frequency, while $v_0 \rightarrow 0$. In that light,
the system to consider analytically must include
the beads $n=-1$, $n=0$ and $n=1$ for the $u$-field only. However,
we were unable to identify an analytical solution of this six-dimensional
nonlinear system (note also that unlike the reduced model considered in Sec.~\ref{sec:reduced} for small $\eps$,
in this case there is no small parameter).
Nevertheless, numerical computations with the single component
system (where there exists a local oscillator at $n=0$)
clearly confirm the validity of the above picture.

\begin{figure}[tbph]
\begin{center}
\includegraphics[width=\textwidth]{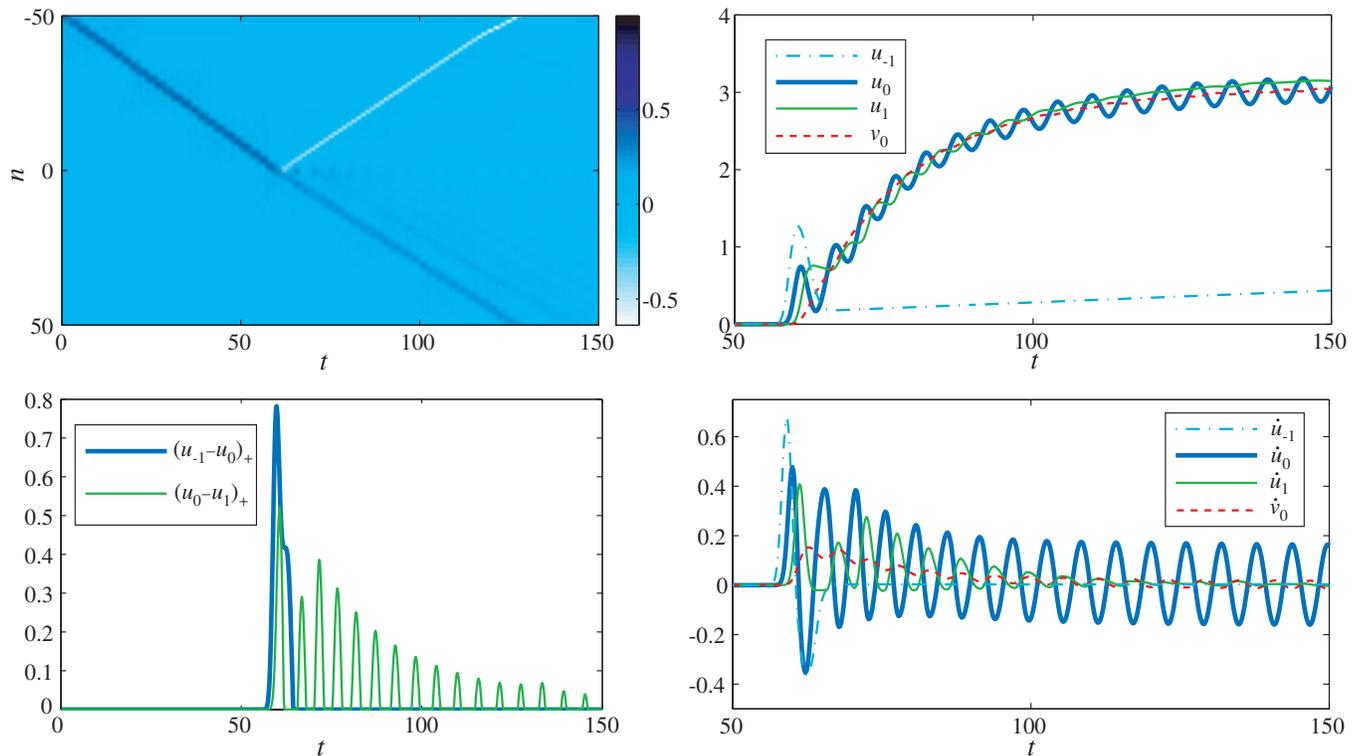}
\caption{The same properties as for the Fig.~\ref{fig01} are illustrated
but now for the case of $\epsilon=10$. Here, it is the lighter
mass $m_1$ executing the fast scale oscillations, while opening a gap
from the site to its left and presenting a cascade of alternating
gap openings and compression events with its right neighbor which, in turn,
lead to the successive emission of progressively smaller amplitude (and thus
slower) waves within the solitary wave train that emerges in the space-time
contour plot.}
\label{fig10}
\end{center}
\end{figure}

We now examine the interaction of the traveling wave with the defect for intermediate values of $\eps$.
We start with the case $\epsilon=10$ shown in Fig.~\ref{fig10}, when the secondary mass
is still considerably larger than the primary one.
First, it should be noted that as expected here, the fast-scale oscillations
are performed by the mass $m_1$ (contrary to what was the case
for small $\epsilon$, where the mass $m_2$ was the one performing
the fast-scale vibrations). Second, the phenomena observed in this case are
significantly different from what we have seen at small $\epsilon$. There still
exists a small fraction of the energy which remains trapped at the
central site. However, the principal phenomenology does not
consist of a single transmitted and reflected wave, as was the
case for small $\epsilon$. In this case, the dominant reflected wave
may be a single one, as was the case for small $\epsilon$ (as shown in Fig.~\ref{fig10}), but there
is a cascade of transmitted waves which are somewhat reminiscent
of the corresponding train of transmitted solitary waves
discussed in~\cite{hascoet} for large mass defects in Hertzian
chains; see also the work of~\cite{sen} for a similar problem where
a solitary wave train also emerges as a consequence of a chain being 
stricken by a heavier bead,
a role that in our system is played by the defect bead.
The top right panel of the figure
illustrates the displacements and allows the observation of the gap
openings where there is no force after the impact of the traveling
wave.
We see that upon the transmission of the original traveling wave,
there is a gap opening between the defect site and its left neighbor ($n=0$ and $n=-1$;
see the thick blue curve in the bottom left panel in Fig.~\ref{fig10}),
which initiates the principal reflected wave traveling to the left.
These two sites never interact again.  However, the interaction of the
$n=0$ defect site
with the site to the right of it ($n=1$) is different. What is observed
is that there is a series of gap openings which are followed (each in
turn) by subsequent compression intervals (see the thin green curve in the bottom left panel).
Each one of these sequences produces a new
traveling wave, as can be seen in the top left panel. However, naturally, as the energy trapped within the central
site keeps decreasing upon the release of subsequent traveling waves,
the amplitude of each later (emerging) wave in the sequence is weaker
than that of its predecessors and hence its speed is also smaller.
This, in turn, justifies this sequence of decreasing amplitude and
speed traveling waves emitted
by the defect.

We note that a similar sequence of gap openings between the beads at
$n=0$ and $n=1$
was also seen at $\eps=0.1$ (see the thin green curve in the inset in the bottom left panel Fig.~\ref{fig01}). However, due to the much smaller amplitude
and higher frequency of oscillations of $u_0(t)$ in that case, the nonzero interaction forces during the compression intervals and
the duration of these intervals were too small to initiate additional traveling waves.

\begin{figure}[tbph]
\begin{center}
\includegraphics[width=\textwidth]{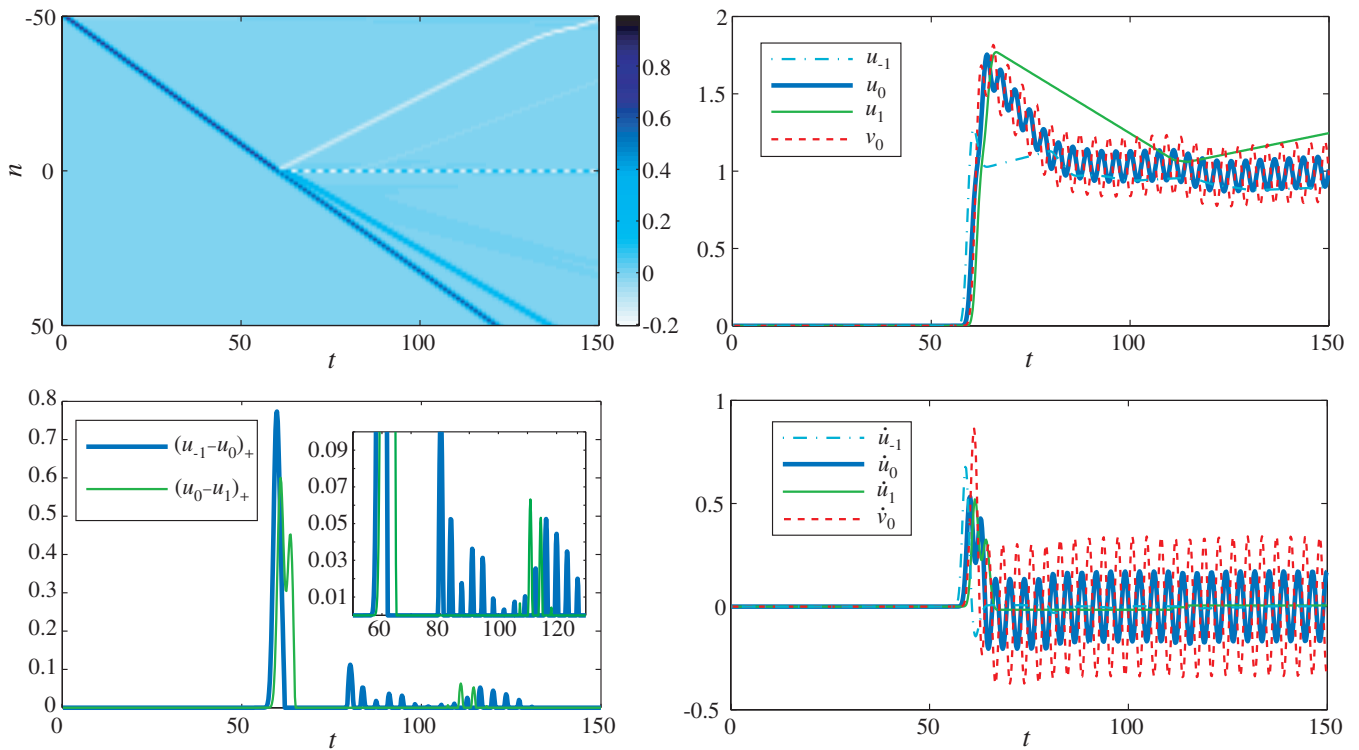}
\caption{The case of $\epsilon=0.5$, close to the value at which the maximal trapping occurs.
The same diagnostics as for $\epsilon=0.1$ in the case of Fig.~\ref{fig01}
are shown. In this case the results reveal vibration over the fast scale of both
$u_0$ and $v_0$ at the defect site and formation of only a second wave
traveling to the right.}
\label{fig05}
\end{center}
\end{figure}

The dynamics for the case when the two masses are comparable
is represented in Fig.~\ref{fig05}, which corresponds to the value $\epsilon=0.5$.
It should be noted here that this case represents the parameter
close to the value at which the trapping fraction of the energy is maximal. In particular, as
observed in Fig.~\ref{line}, there is a clearly defined maximum
in the fraction of energy that can be trapped by the system;
this is another unique feature of our system. Not only can energy
be trapped by the mass-with-mass defect but the trapping has
a non-monotonic dependence on the ratio of the masses. As illustrated
in the figure, here the scenario is different than the ones
observed before. Here, the defect bead does not
lead to a gap opening with respect to the bead to its left, but
rather (predominantly) with respect to the bead to its right. In this case, 
we observe a two-peaked structure in $(u_1-u_0)_+$
(see the bottom left panel of Fig.~\ref{fig05}) which
appears to nucleate two traveling waves moving in the right half
of the chain. While a gap indeed opens past this principal
interaction between $n=0$ and $n=1$, the beads do interact anew
at a much later time (around $t=110$). Yet, similarly to the interactions
shown in Fig.~\ref{fig01}, such later (and considerably weaker)
interactions do not produce appreciable
solitary waves traveling to the right of $n=0$.

\begin{figure}[tbph]
\begin{center}
\includegraphics[width=\textwidth]{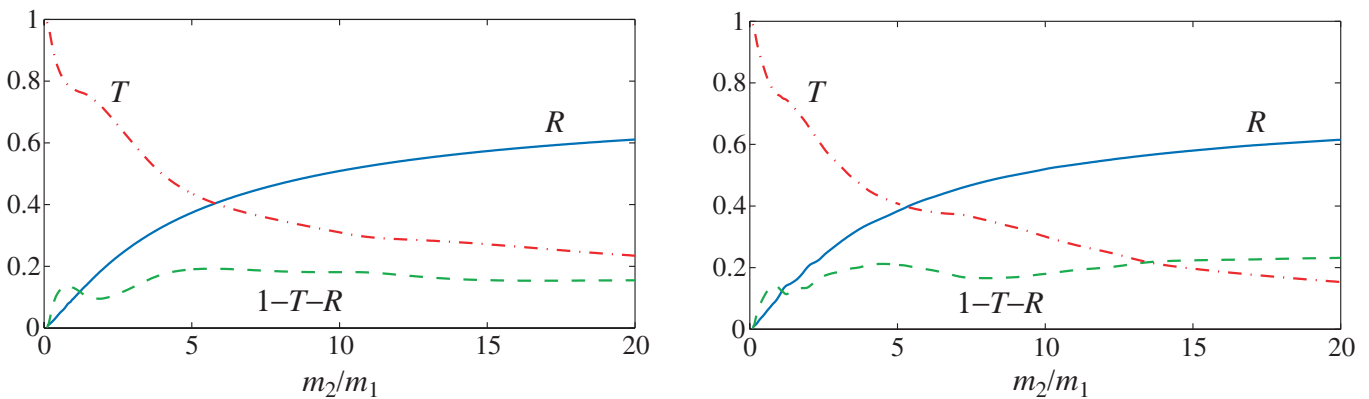}
\end{center}
\caption{The same diagnostics as those of Fig.~\ref{line} are
shown here (transmitted $T$, reflected $R$ and trapped $1-T-R$
fractions of the energy) as a function of the mass ratio, but now
for the case of two adjacent mass defects (left) and that of three 
adjacent mass defects (right).}
\label{addition}
\end{figure}

Finally, we turn to a setting where there exist additional defects
within the chain. This is shown in Fig.~\ref{addition}, which contains
both the case of two adjacent MwM defects (left panel), as well as that with
three such adjacent defects (right panel).  It can be clearly discerned that
the addition of further defects significantly enhances the trapped
fraction of the original
energy within the defect sites. This fraction increases
by about $10$\% (for comparable values of $m_2/m_1$), when the defective
region expands from one to two sites, and a comparable increase is observed
when a third site is added. It is observed that this increase is
at the expense of the transmitted fraction,
which is well below $20$\% ($\approx 15$\% and well below the trapped
fraction) for the three-MwM defect case. Interestingly, the trapped
fraction preserves its oscillatory structure (notice that a less
pronounced, yet somewhat similar variation is observed in the
transmitted fraction), but the peak locations vary, as the number
of defect sites is increased; e.g. the first, most pronounced peak
occurs at larger values of $m_2/m_1$.

\section{Analysis of the small mass ratio case}
\label{sec:reduced}
We now present a theoretical formulation which enables a semi-quantitative
understanding of the situation at hand for the case of small $\epsilon$.
We will use two different approaches, one of which involves perturbation analysis of the full system
and the other is based on multiscale analysis of a reduced system.

\subsection{Perturbation analysis}
Assume that there exists a traveling wave satisfying the standard Hertzian chain equation
$\ddot{u}_n = (u_{n-1}-u_n)^{3/2}_+ - (u_n - u_{n+1})^{3/2}_+$, where
the solution can be well approximated by the Nesterenko solitary wave \cite{nesterenko1}
with
\beq
\dot{u}_n=\begin{cases}
Ac\sin^4\biggl(\sqrt{\dfrac{2}{5}} (n-c(t-t_0))\biggr), & t_0+\dfrac{n-\sqrt{\frac{5}{2}}\pi}{c} \leq t \leq t_0+\dfrac{n}{c},\\
0, & \text{otherwise}
\end{cases}
\label{eq:TW}
\eeq
where $c^2=(4/5) A^{1/2}$ and $t_0$
is the time when the wave leaves the $n=0$ site.
Now, focusing on the difference $w_0=u_0-v_0$, we note
that it satisfies the equation
\beq
\ddot{w}_0 + \omega^2 w_0= (u_{-1}-u_0)^{3/2}_+ -
(u_0 - u_{1})^{3/2}_+,
\label{approxim}
\eeq
where
\[
\omega=\sqrt{1+\epsilon^{-1}}
\]
is the natural frequency of the MwM defect. While Eq.(~\ref{approxim}) is exact, observe that to a leading order its right hand side can be approximated by $\ddot{u}_0$. This means that the linear oscillator is driven by the propagating traveling wave, which we further approximate by the above Nesterenko expression \eqref{eq:TW} at $n=0$, with $\ddot{u}_0$ obtained by differentiating \eqref{eq:TW} once with respect to time.
Notice that this drive is only active within the time interval
$[t_0-(\pi\sqrt{5/2}/c),t_0]$, i.e. as the wave passes over $n=0$ site.

Within this approximation, Eq.~(\ref{approxim}) can be solved exactly
to give
\beq
w_0 = \int_0^t \frac{\sin(\omega (t-\tau))}{\omega} \ddot{u}_0(\tau) d \tau.
\label{approxim2}
\eeq
This simplified model has the advantage that it can be evaluated explicitly:
\beq
w_0=\begin{cases} 0, & t<t_0-\frac{\pi\sqrt{5}}{\sqrt{2}c}\\
\frac{A c^2}{D}\biggl(\sqrt{10}(5 \omega^2-32 c^2) \sin(2\phi)
+ \sqrt{\frac{5}{2}} (8 c^2-5 \omega^2) \sin(4 \phi)
+ \frac{96 c^3}{\omega} \sin(\omega (\frac{\pi\sqrt{5}}{\sqrt{2}c}+t-t_0))\biggr), &
t_0-\frac{\pi\sqrt{5}}{\sqrt{2}c} \leq t \leq t_0\\
-\frac{192 A c^5}{D\omega}\sin(\omega\frac{\pi\sqrt{5}}{2\sqrt{2}c})\cos(\omega(t-t_0+\frac{\pi\sqrt{5}}{2\sqrt{2}c})), & t>t_0,
\end{cases}
\label{approxim3}
\eeq
where $\phi=\sqrt{2/5}(c(t-t_0))$ and
$D=256 c^4  - 200 c^2 \omega^2 +25 \omega^4$.
Note that it corroborates the numerical observation that
the trapped oscillation of the mass-with-mass defect is executed
with the natural frequency $\omega$ of the
relevant oscillator.

\begin{figure}[tbph]
\begin{center}
\includegraphics[width=\textwidth]{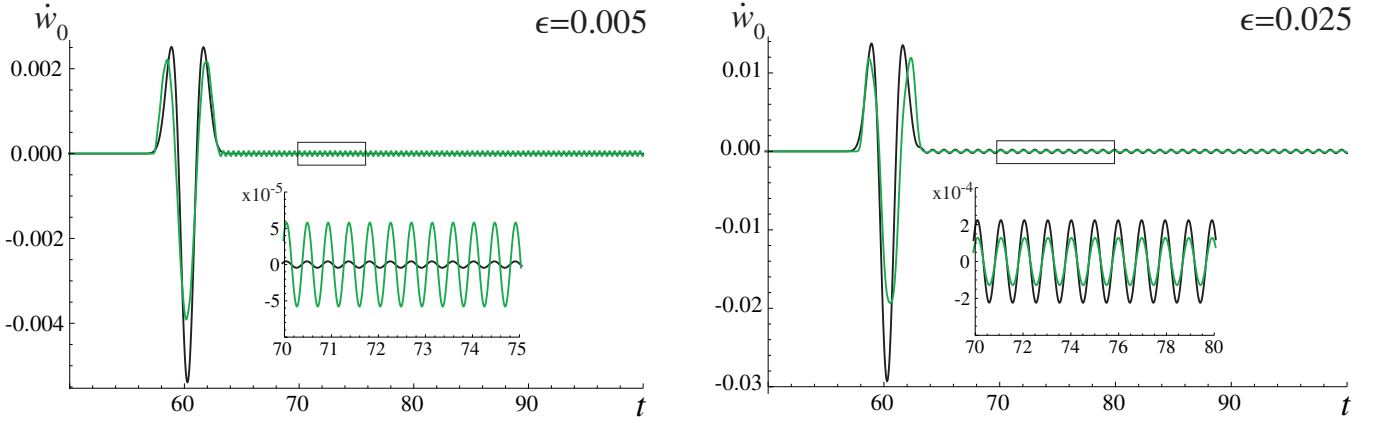}
\caption{The time derivative of the difference $w_0(t)=u_0(t)-v_0(t)$ given by
the direct numerical computation (black) and
the analytical approximation (green) for $\eps=0.005$ and $\eps=0.025$. The insets zoom in on the regions inside the rectangles.}
\label{line1}
\end{center}
\end{figure}

The results of this approximation for small values of $\epsilon$ are
illustrated in the example of Fig.~\ref{line1}.
Here the traveling wave is initiated
by the standard initial condition above and its resulting
speed $c$ and amplitude $A$ are computed from the simulation prior to
its impact with the mass-with-mass defect (we obtain $c \approx 0.841$ and $A \approx 0.811$).
Then the time derivative of the expression
of Eq.~(\ref{approxim3}) (shown by the
green line) is evaluated
and compared, after an appropriate time shift $t_0$, with the direct result of the numerical
simulation (black line) at the central site. It can be seen
that the analytical solution captures the
numerical result quite well
qualitatively and, as expected, properly captures the frequency of the residual vibration
trapped at the central site. It does not, however, capture its amplitude in general, which can be either larger (see, for example, the 
discrepancy at $\eps=0.005$) or smaller (e.g. $\eps=0.025$) than the numerical value. It also
underestimates the amplitude of the large pulse that precedes it. This happens because
our approximation neglects the effect the defect has on the propagating wave which adjusts its speed and shape prior to reaching the
$n=0$ site.

\subsection{Small $\eps$ limit: two-bead problem}

To better understand the small $\eps$ limit, we now
turn to a simplification of the original problem in which we only
consider the system involving two beads. Neglecting the left part
of the chain can be justified by the fact that the numerical simulations show that after the defect bead has been
kicked by the previous one, there is a gap opening between the two and therefore they
do not interact again.
Hence, the simplest configuration that could capture the transmission of
the wave from $n=0$ to $n=1$, as well as the trapping of the energy
in the $n=0$ site would be the two-site setting examined below.
Importantly, these two phenomena exhibit a separation of time scales:
the oscillation within the defect bead is one of a fast time scale,
while the interaction between $n=0$ and $n=1$ occurs in a slower
time scale, enabling the multiscale analysis presented below.

\begin{figure}[tbph]
\begin{center}
\includegraphics[width=5.in]{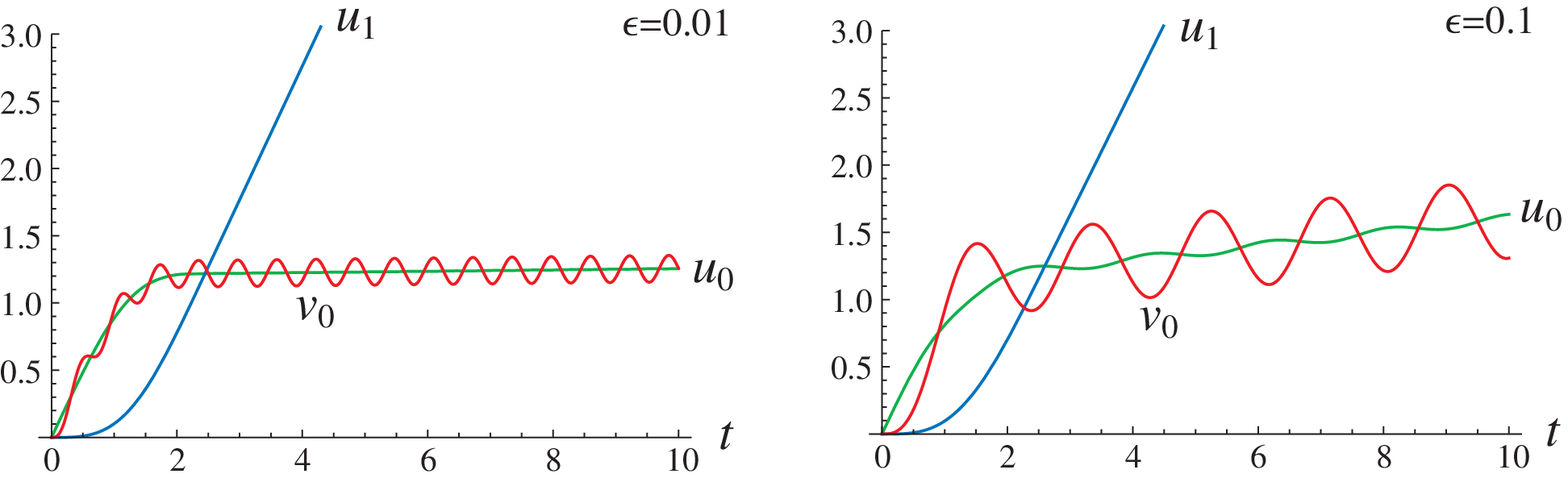}
\caption{Numerical solution of \eqref{eq:two-bead}, \eqref{eq:ICs} at $\eps=0.01$ and $\eps=0.1$.}
\label{fig:numer_soln}
\end{center}
\end{figure}

The simplified dynamical equations then read:
\beq
\begin{split}
\ddot{u}_0&=-(u_0-u_1)^{3/2}_+-(u_0-v_0)\\
\eps \ddot{v}_0&=u_0-v_0\\
\ddot{u}_1&=(u_0-u_1)^{3/2}_+.
\end{split}
\label{eq:two-bead}
\eeq
In considering this reduced model, we follow the approach in \cite{vakakis} for a problem with a light mass defect, with the goal to
qualitatively capture the dynamics of the system during an initial time period after the propagating wave reaches the mass-with-mass defect.
For example, consider the initial conditions (emulating the kick provided
through the gap-opening interaction with $n=-1$)
\beq
u_0(0)=v_0(0)=u_1(0)=\dot{v}_0(0)=\dot{u}_1(0)=0, \quad \dot{u}_0(0)=1.
\label{eq:ICs}
\eeq
The numerical solution of the system \eqref{eq:two-bead} at $\eps=0.01$ and $\eps=0.1$ subject to \eqref{eq:ICs} is shown in Fig.~\ref{fig:numer_soln}.
By numerically integrating the simplified system and comparing the results with the previously obtained ones for the full system, we see that
the simplified model qualitatively captures the behavior of the full system at small $\eps$ in that $v_0(t)$ and $u_0(t)$ both oscillate, $v_0(t)$ with
a larger amplitude, about some
average value that increases with time, initially rapidly and then more slowly.

To analytically approximate \eqref{eq:two-bead} at small $\eps$, we use a
two-timing approach~\cite{vakakis}. We introduce the fast
time $\tau=t/\sqrt{\eps}$. This rescaling of time yields:
\beq
\begin{split}
\dfrac{d^2u_0}{d\tau^2}&=-\eps(u_0-u_1)^{3/2}_+-\eps(u_0-v_0)\\
\dfrac{d^2v_0}{d\tau^2}&=u_0-v_0\\
\dfrac{d^2u_1}{d\tau^2}&=\eps(u_0-u_1)^{3/2}_+.
\end{split}
\label{eq:two-bead_fast}
\eeq
Substituting the expansion
\[
\begin{split}
u_0(\tau)&=U_0^{(0)}(\tau,t)+\sqrt{\eps}U_0^{(1)}(\tau,t)+\eps U_0^{(2)}(\tau,t)+\eps^{3/2}U_0^{(3)}(\tau,t)+O(\eps^2)\\
v_0(\tau)&=V_0^{(0)}(\tau,t)+\sqrt{\eps}V_0^{(1)}(\tau,t)+\eps V_0^{(2)}(\tau,t)+\eps^{3/2}V_0^{(3)}(\tau,t)+O(\eps^2)\\
u_1(\tau)&=U_1^{(0)}(\tau,t)+\sqrt{\eps}U_1^{(1)}(\tau,t)+\eps U_1^{(2)}(\tau,t)+\eps^{3/2}U_1^{(3)}(\tau,t)+O(\eps^2)
\end{split}
\]
in \eqref{eq:two-bead_fast}, expanding the right hand side in terms of $\eps$ and
keeping the terms up to $O(\eps^2$), we obtain
\[
\begin{split}
\dfrac{\partial^2 U_0^{(0)}}{\partial \tau^2}&+\sqrt{\eps}\left(\dfrac{\partial^2 U_0^{(1)}}{\partial\tau^2}
+ 2\dfrac{\partial^2 U_0^{(0)}}{\partial\tau\partial t} \right)
+\eps\left(\dfrac{\partial^2 U_0^{(0)}}{\partial t^2}+2\dfrac{\partial^2 U_0^{(1)}}{\partial t\partial\tau}
+\dfrac{\partial^2 U_0^{(2)}}{\partial \tau^2}\right)+\eps^{3/2}\left(\dfrac{\partial^2 U_0^{(1)}}{\partial t^2}
+2\dfrac{\partial^2 U_0^{(2)}}{\partial t\partial\tau}+\dfrac{\partial^2 U_0^{(3)}}{\partial \tau^2}\right)\\
&=
-\eps\left\{\left(U_0^{(0)}-U_1^{(0)}\right)_+^{3/2}+U_0^{(0)}-V_0^{(0)}\right\}
-\eps^{3/2}\left\{\dfrac{3}{2}\left(U_0^{(0)}-U_1^{(0)}\right)_+^{1/2}\left(U_0^{(1)}-U_1^{(1)}\right)+U_0^{(1)}-V_0^{(1)}\right\}
\end{split}
\]
\[
\begin{split}
\dfrac{\partial^2 V_0^{(0)}}{\partial \tau^2}&+\sqrt{\eps}\left(\dfrac{\partial^2 V_0^{(1)}}{\partial\tau^2}
+ 2\dfrac{\partial^2 V_0^{(0)}}{\partial\tau\partial t} \right)
+\eps\left(\dfrac{\partial^2 V_0^{(0)}}{\partial t^2}+2\dfrac{\partial^2 V_0^{(1)}}{\partial t\partial\tau}
+\dfrac{\partial^2 V_0^{(2)}}{\partial \tau^2}\right)+\eps^{3/2}\left(\dfrac{\partial^2 V_0^{(1)}}{\partial t^2}
+2\dfrac{\partial^2 V_0^{(2)}}{\partial t\partial\tau}+\dfrac{\partial^2 V_0^{(3)}}{\partial \tau^2}\right)\\
&=U_0^{(0)}-V_0^{(0)}+\sqrt{\eps}\left(U_0^{(1)}-V_0^{(1)}\right)+\eps\left(U_0^{(2)}-V_0^{(2)}\right)+\eps^{3/2}\left(U_0^{(3)}-V_0^{(3)}\right)
\end{split}
\]
\[
\begin{split}
\dfrac{\partial^2 U_1^{(0)}}{\partial \tau^2}&+\sqrt{\eps}\left(\dfrac{\partial^2 U_1^{(1)}}{\partial\tau^2}
+ 2\dfrac{\partial^2 U_1^{(0)}}{\partial\tau\partial t} \right)
+\eps\left(\dfrac{\partial^2 U_1^{(0)}}{\partial t^2}+2\dfrac{\partial^2 U_1^{(1)}}{\partial t\partial\tau}
+\dfrac{\partial^2 U_1^{(2)}}{\partial \tau^2}\right)+\eps^{3/2}\left(\dfrac{\partial^2 U_1^{(1)}}{\partial t^2}
+2\dfrac{\partial^2 U_1^{(2)}}{\partial t\partial\tau}+\dfrac{\partial^2 U_1^{(3)}}{\partial \tau^2}\right)\\
&=
\eps\left(U_0^{(0)}-U_1^{(0)}\right)_+^{3/2}
+\dfrac{3}{2}\eps^{3/2}\left(U_0^{(0)}-U_1^{(0)}\right)_+^{1/2}\left(U_0^{(1)}-U_1^{(1)}\right)
\end{split}
\]
Then $O(1)$ problem reads
\beq
\dfrac{\partial^2 U_0^{(0)}}{\partial \tau^2}=0, \quad \dfrac{\partial^2 U_1^{(0)}}{\partial \tau^2}=0, \quad
\dfrac{\partial^2 V_0^{(0)}}{\partial \tau^2}=U_0^{(0)}-V_0^{(0)}
\label{eq:prob1}
\eeq
Solving the above equations and eliminating the secular terms (which feature unbounded growth in $\tau$), we obtain
\beq
U_0^{(0)}=B_0(t), \quad U_1^{(0)}=B_1(t)
\label{eq:U0}
\eeq
and a fast oscillation of the defect as
\beq
V_0^{(0)}=C(t)\cos\tau+D(t)\sin\tau+B_0(t).
\label{eq:V0_1}
\eeq
Next we consider the $O(\sqrt{\eps})$ problem:
\beq
\dfrac{\partial^2 U_0^{(1)}}{\partial \tau^2}+2\dfrac{\partial^2 U_0^{(0)}}{\partial\tau\partial t}=0, \quad
\dfrac{\partial^2 U_1^{(1)}}{\partial \tau^2}+2\dfrac{\partial^2 U_1^{(0)}}{\partial\tau\partial t}=0, \quad
\dfrac{\partial^2 V_0^{(1)}}{\partial \tau^2}+2\dfrac{\partial^2 V_0^{(0)}}{\partial\tau\partial t}=U_0^{(1)}-V_0^{(1)}
\label{eq:prob2}
\eeq
From \eqref{eq:U0}, it follows that
\[
\dfrac{\partial^2 U_0^{(0)}}{\partial\tau\partial t}=\dfrac{\partial^2 U_1^{(0)}}{\partial\tau\partial t}=0,
\]
and hence the first two equations in \eqref{eq:prob2} imply that
\beq
U_0^{(1)}=E_0(t), \quad U_1^{(1)}=E_1(t).
\label{eq:U1}
\eeq
Differentiating \eqref{eq:V0_1} to find the mixed derivative in the third equation in \eqref{eq:prob2},
\[
\dfrac{\partial^2 V_0^{(0)}}{\partial\tau\partial t}=-C'(t)\sin\tau+D'(t)\cos\tau,
\]
we find that we must have $C'(t)=D'(t)=0$ in order to avoid secular terms in $V_0^{(1)}$. Thus $C$ and $D$ are constant, and we have
\beq
V_0^{(0)}=C\cos\tau+D\sin\tau+B_0(t),
\label{eq:V0}
\eeq
while the last equation in \eqref{eq:prob2} and \eqref{eq:U1} yield
\beq
V_0^{(1)}=F(t)\cos\tau+G(t)\sin\tau+E_0(t).
\label{eq:V1_1}
\eeq
In the subsequent order, namely $O(\eps)$, we have
\beq
\begin{split}
&\dfrac{\partial^2 U_0^{(0)}}{\partial t^2}+2\dfrac{\partial^2 U_0^{(1)}}{\partial t\partial\tau}
+\dfrac{\partial^2 U_0^{(2)}}{\partial \tau^2}=-\left(U_0^{(0)}-U_1^{(0)}\right)_+^{3/2}-\left(U_0^{(0)}-V_0^{(0)}\right)\\
&\dfrac{\partial^2 V_0^{(0)}}{\partial t^2}+2\dfrac{\partial^2 V_0^{(1)}}{\partial t\partial\tau}
+\dfrac{\partial^2 V_0^{(2)}}{\partial \tau^2}=U_0^{(2)}-V_0^{(2)}\\
&\dfrac{\partial^2 U_1^{(0)}}{\partial t^2}+2\dfrac{\partial^2 U_1^{(1)}}{\partial t\partial\tau}
+\dfrac{\partial^2 U_1^{(2)}}{\partial \tau^2}=\left(U_0^{(0)}-U_1^{(0)}\right)_+^{3/2}
\end{split}
\label{eq:prob3}
\eeq
Using \eqref{eq:U0}, \eqref{eq:U1} and recalling the third equation in \eqref{eq:prob1}, we can rewrite
the first equation in \eqref{eq:prob3} as
\[
\dfrac{\partial^2}{\partial \tau^2}\left(U_0^{(2)}+V_0^{(0)}\right)=
-B_0''(t)-(B_0(t)-B_1(t))_+^{3/2}.
\]
To eliminate secular terms from $U_0^{(2)}$, we must then set the right hand side of this equation to zero:
\beq
B_0''(t)=-(B_0(t)-B_1(t))_+^{3/2}.
\label{eq:B0_ODE}
\eeq
Elimination of the linear term in $\tau$ then means that $U_0^{(2)}+V_0^{(0)}$ is a function of $t$ only, yielding
\beq
U_0^{(2)}=H_0(t)-C\cos\tau-D\sin\tau,
\label{eq:U2_1}
\eeq
where we used \eqref{eq:V0}.
Similarly, using \eqref{eq:U0} and \eqref{eq:U1} in the
last equation of \eqref{eq:prob3} and
eliminating  the secular terms in $U_1^{(2)}$, we obtain
\beq
B_1''(t)=(B_0(t)-B_1(t))_+^{3/2}
\label{eq:B1_ODE}
\eeq
and
\beq
U_1^{(2)}=H_1(t).
\label{eq:U2_2}
\eeq
Considering now the second equation in \eqref{eq:prob3} and substituting the mixed derivative of \eqref{eq:V1_1} and \eqref{eq:U2_1}, we have
\[
\dfrac{\partial^2 V_0^{(2)}}{\partial \tau^2}+V_0^{(2)}=-(2G'(t)+C)\cos\tau+(2F'(t)-D)\sin\tau+H_0(t)-B_0''(t).
\]
To avoid secular terms, we must set the coefficients in front of $\sin\tau$ and $\cos\tau$ in the right hand side to zero, which yields
$F(t)=(D/2)t+D_0$ and $G(t)=-(C/2)t+C_0$, where $C_0$ and $D_0$ are constant. Thus we have
\beq
V_0^{(1)}=\left(\dfrac{D}{2}t+D_0\right)\cos\tau+\left(C_0-\dfrac{C}{2}t\right)\sin\tau+E_0(t),
\label{eq:V1}
\eeq
while
\[
V_0^{(2)}=I(t)\cos\tau+J(t)\sin\tau+H_0(t)-B_0''(t).
\]
Finally, we turn to the $O\left(\eps^{3/2}\right)$ problem:
\beq
\begin{split}
&\dfrac{\partial^2 U_0^{(1)}}{\partial t^2}+2\dfrac{\partial^2 U_0^{(2)}}{\partial t\partial\tau}
+\dfrac{\partial^2 U_0^{(3)}}{\partial \tau^2}=-\dfrac{3}{2}\left(U_0^{(0)}-U_1^{(0)}\right)_+^{1/2}\left(U_0^{(1)}-U_1^{(1)}\right)
-\left(U_0^{(1)}-V_0^{(1)}\right)\\
&\dfrac{\partial^2 V_0^{(1)}}{\partial t^2}+2\dfrac{\partial^2 V_0^{(2)}}{\partial t\partial\tau}
+\dfrac{\partial^2 V_0^{(3)}}{\partial \tau^2}=U_0^{(3)}-V_0^{(3)}\\
&\dfrac{\partial^2 U_1^{(1)}}{\partial t^2}+2\dfrac{\partial^2 U_1^{(2)}}{\partial t\partial\tau}
+\dfrac{\partial^2 U_1^{(3)}}{\partial \tau^2}=\dfrac{3}{2}\left(U_0^{(0)}-U_1^{(0)}\right)_+^{1/2}\left(U_0^{(1)}-U_1^{(1)}\right)
\end{split}
\label{eq:prob4}
\eeq
Proceeding as in the $O(\eps)$ case, we use \eqref{eq:U1}, \eqref{eq:U2_1}, the third equation in \eqref{eq:prob2} and \eqref{eq:V0} to rewrite
the first equation in \eqref{eq:prob4} as
\[
\dfrac{\partial^2}{\partial \tau^2}\left(U_0^{(3)}+V_0^{(1)}\right)=
-E_0''(t)-\dfrac{3}{2}\left(B_0(t)-B_1(t)\right)_+^{1/2}(E_0(t)-E_1(t)).
\]
To eliminate secular terms from $U_0^{(3)}$, we must then set the right hand side of this equation to zero:
\beq
E_0''(t)=-\dfrac{3}{2}\left(B_0(t)-B_1(t)\right)_+^{1/2}(E_0(t)-E_1(t)).
\label{eq:E0_ODE}
\eeq
Then, again eliminating the secular terms, we get that $U_0^{(3)}+V_0^{(1)}$ is only a function of $t$, which in light of \eqref{eq:V1} yields
\[
U_0^{(3)}=K_0(t)-\left(\dfrac{D}{2}t+D_0\right)\cos\tau-\left(C_0-\dfrac{C}{2}t\right)\sin\tau.
\]
Similarly, \eqref{eq:U1}, \eqref{eq:U2_2} and the last equation in \eqref{eq:prob4} result in
\beq
E_1''(t)=\dfrac{3}{2}\left(B_0(t)-B_1(t)\right)_+^{1/2}(E_0(t)-E_1(t))
\label{eq:E1_ODE}
\eeq
and $U_1^{(3)}=K_1(t)$.

So up to order $O(\eps)$ we have
\[
\begin{split}
u_0(t)&=B_0(t)+\sqrt{\eps}E_0(t), \quad u_1(t)=B_1(t)+\sqrt{\eps}E_1(t), \\
v_0(t)&=B_0(t)+C\cos\dfrac{t}{\sqrt{\eps}}+D\sin\dfrac{t}{\sqrt{\eps}}+\sqrt{\eps}\left\{E_0(t)+\left(D_0+\dfrac{D}{2}t\right)\cos\dfrac{t}{\sqrt{\eps}}
+\left(C_0-\dfrac{C}{2}t\right)\sin\dfrac{t}{\sqrt{\eps}}\right\},
\end{split}
\]
where $B_0(t)$, $B_1(t)$ are found by solving \eqref{eq:B0_ODE}, \eqref{eq:B1_ODE}, $E_0(t)$, $E_1(t)$ satisfy \eqref{eq:E0_ODE}, \eqref{eq:E1_ODE}, and
the initial conditions for these functions, as well as the constants $D_0$, $D$, $C_0$, $C$,
are found from the initial conditions for $u_0(t)$, $u_1(t)$ and $v_0(t)$.

Now we consider the initial conditions \eqref{eq:ICs}. Then $u_0(0)=v_0(0)=u_1(0)=0$, which imply that $B_0(0)=B_1(0)=E_0(0)=E_1(0)=0$
and $C=D_0=0$. The initial condition $\dot{u}_0(0)=1$ implies $B_0'(0)=1$ and $E_0'(0)=0$, while $\dot{u}_1(0)=0$ implies $B_1'(0)=E_1'(0)=0$.
Differentiating $v_0(t)$ and using $\dot{v}_0(0)=0$ and the above results, we get
\[
\dot{v}_0(0)=1+\dfrac{D}{\sqrt{\eps}}+C_0+\dfrac{D\sqrt{\eps}}{2}=0,
\]
which implies $D=0$ and $C_0=-1$. Since $E_0(t)$, $E_1(t)$ satisfy the linear system \eqref{eq:E0_ODE}, \eqref{eq:E1_ODE} subject to
the zero initial conditions, they must vanish. Thus we have
\beq
u_0(t)=B_0(t), \quad u_1(t)=B_1(t), \quad
v_0(t)=B_0(t)-\sqrt{\eps}\sin\dfrac{t}{\sqrt{\eps}}
\label{eq:approx1}
\eeq
where $B_0(t)$, $B_1(t)$ satisfy \eqref{eq:B0_ODE}, \eqref{eq:B1_ODE} and the initial conditions $B_0(0)=B_1(0)=B_1'(0)=0$ and $B_0'(0)=1$. To solve
this problem, it is convenient
to introduce the new variables $y_0(t)=(B_0(t)-B_1(t))/2$ and $z_0(t)=(B_0(t)+B_1(t))/2$. Then $z_0(t)$
satisfies $z_0''(t)=0$, $z_0(0)=0$, $z_0'(0)=1/2$, so we have $z(t)=t/2$. Meanwhile, $y_0(t)$ solves
\beq
y_0''+(2y_0)_+^{3/2}=0, \quad y_0(0)=0, \quad y_0'(0)=1/2.
\label{eq:w0_eqn}
\eeq
When $y_0>0$, it satisfies $y_0''+(2y_0)^{3/2}=0$, which together with the initial conditions implies that it lies on the trajectory
\[
(y_0')^2+\dfrac{8}{5}\sqrt{2}y_0^{5/2}=\dfrac{1}{4}
\]
This can be solved for $t$ as a function of $y_0$ in terms of hypergeometric functions. We obtain
\beq
t=\begin{cases} 2y_0\,{}_2F_1[\frac{2}{5},\frac{1}{2},\frac{7}{5},\frac{32}{5}\sqrt{2}y_0^{5/2}], & 0 \leq t \leq t_*\\
             2(t_*-y_0\,{}_2F_1[\frac{2}{5},\frac{1}{2},\frac{7}{5},\frac{32}{5}\sqrt{2}y_0^{5/2}]), & t_* \leq t \leq 2t_*,
  \end{cases}
\label{eq:w0_1}
\eeq
where $t_*=2y_0\,{}_2F_1[\frac{2}{5},\frac{1}{2},\frac{7}{5},1]$ is such that $y_0(t_*)=y_0^*=\left(\dfrac{5}{32\sqrt{2}}\right)^{2/5}$
For $t>2t_*$ we have $y_0''=0$, and since $y_0'(2t_*)=-1/2$ and $y_0(2t_*)=0$, this yields
\beq
y_0(t)=t_*-t/2 \quad \text{for $t>2t_*$}.
\label{eq:w0_2}
\eeq
Recalling that $B_0(t)=y_0(t)+z_0(t)$  and $B_1(t)=z_0(t)-y_0(t)$, we obtain
\beq
u_0(t)=\dfrac{t}{2}+y_0(t), \quad u_1(t)=\dfrac{t}{2}-y_0(t), \quad
v_0(t)=\dfrac{t}{2}+y_0(t)-\sqrt{\eps}\sin\dfrac{t}{\sqrt{\eps}}
\label{eq:approx}
\eeq
where $y_0(t)$ is given by \eqref{eq:w0_1}, \eqref{eq:w0_2}. In particular, for $t \geq 2t_*$ we have $u_0$ becomes constant, $u_0(t)=t_*$, while $u_1$
linearly increases, $u_1(t)=t_*+t/2$.

\begin{figure}[tbph]
\begin{center}
\includegraphics[width=3.0in]{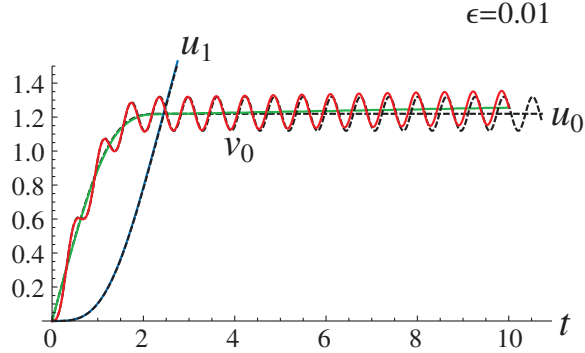}
\caption{Comparison of the numerical solution of \eqref{eq:two-bead}, \eqref{eq:ICs} at $\eps=0.01$ (colored solid curves) and the two-timing approximation up to
$O(\eps)$ (black dashed curves).}
\label{fig:approx}
\end{center}
\end{figure}

The comparison of the approximation \eqref{eq:approx} (black curves) and the numerical solution (colored curves) is
shown in Fig.~\ref{fig:approx} at $\eps=0.01$.
One can see that the approximation works very well for $t<5$, but starts deviating for larger $t$ because it does not capture the slight increase
in $u_0(t)$ for $t>2t_*$. It also does not capture the small-amplitude oscillations of $u_0(t)$, which become visible at larger $\eps$; see, for example,
Fig.~\ref{fig:numer_soln} at $\eps=0.1$. To capture these effects, one needs to include higher-order terms. Note, however, that the function $(u_0-u_1)^{3/2}_+$ is not smooth at zero
and cannot be expanded beyond the first derivative term at this point.
Hence, while this approach is valuable in analytical understanding
of the leading order dynamics in the simplified two-site system,
it also has its limitations with respect to some of the higher
order effects therein.

Nevertheless, the analytical approximations considered in this section
offer a detailed quantitative picture of the energy trapping and residual oscillatory
dynamics within the defect (and capture its frequency), as well as of the detailed exchange
dynamics between sites $n=0$ and $n=1$.

\section{Conclusions and Future Challenges}
\label{sec:conclusions}

In the present work, we have explored the
propagation of waves in a granular chain in which a mass-with-mass
defect is present. In this setting, a local oscillator
with additional parameters arises, the most significant of which
is the ratio between the mass-with-mass defect and that of the
rest of the masses within the chain.  We have considered the problem
as a function of this parameter numerically and wherever
possible also analytically.

We found that in the case of the small defect to bead mass ratio,
the traveling wave remains essentially unaltered, except for the fact
that a part of its energy is reflected and a part of its energy is trapped in the form of localized oscillation.
The trapping of energy is an exclusive feature of this system that was not observed in chains
containing mass defects. This phenomenon was studied analytically
in two distinct ways. One of them involved a direct perturbative approach
based on the fact that the local oscillator is principally driven
by the weakly affected (in this case) solitary wave. The second
one was a multiscale technique applied on a reduced, two-beads system based on two-timing which revealed
the key role of the dynamics/interaction of the defect site $n=0$
with the following one ($n=1$). We also studied the effect of the defect for all the representative values of the mass ratio
using numerical methods. In the limit of very large
mass ratio, we found that the reflection is more significant than the transmission
and a considerable amount of trapping still occurs. In the intermediate
mass ratio cases we found the potential of exciting multiple waves
either one directly after the other, or through a sequence of gap openings
leading to a train of solitary waves emitted towards the right of
the defect. In each case, we studied the dynamics of the interaction with the defect
by computing the trajectories of the beads in the vicinity of the defect ($n=-1$, $n=0$ and $n=1$).

The present study suggests
many possible future investigations on systems containing MwM defects. On one hand,
the large mass ratio limit
would certainly be worthwhile to consider analytically. We also found that the
energy trapped in the MwM defect shows a non-monotonic dependence on
the mass ratio which has not yet been explained. And finally,
this study could be extended to cover the propagation of traveling waves in ordered systems
where all the beads have a MwM defect.
These topics are currently under consideration and will be reported
in future publications.

\vspace{2mm}

{\it Acknowledgements.} CD acknowledges support
from the National Science Foundation, grant number CMMI-844540 (CAREER) and NSF
1138702. PGK acknowledges support from the US National 
Science Foundation
under grant CMMI-1000337, the US Air Force under grant FA9550-12-1-0332, 
the Alexander von Humboldt Foundation, as well as the Alexander S. Onassis 
Public Benefit Foundation.
The work of AV was supported by the NSF grant DMS-1007908.











\end{document}